\documentclass[conference]{IEEEtran} 
\usepackage{amsmath}
\usepackage{amssymb}
\usepackage{tabularx}
\usepackage{mathtools}
\usepackage{graphicx}
\usepackage{optidef}
\usepackage{amsfonts}
\usepackage{algorithm}
\usepackage{algorithmic}
\usepackage{caption}
\usepackage{subcaption}
\usepackage{amsthm}

\usepackage{xcolor}
\usepackage{breqn}
\usepackage{cuted}
\usepackage{balance}
\usepackage[top=0.75in, left=0.64in, right=0.64in, bottom=1.04in, letterpaper]{geometry}

\ifCLASSINFOpdf
\else
\fi
\DeclareUnicodeCharacter{2212}{-}

\begin{document}
%
\title{Towards Resilient 6G O-RAN: An Energy-Efficient URLLC Resource Allocation Framework}
%
%
%
\author{
     Rana M. Sohaib\IEEEauthorrefmark{2}, \IEEEauthorblockN {Syed Tariq Shah\IEEEauthorrefmark{1}, 
     Poonam Yadav\IEEEauthorrefmark{2}} 
    \IEEEauthorblockA{
\IEEEauthorblockA{\IEEEauthorrefmark{2}Department of CS, University of York, York, UK\\ Emails: \{rana.sohaib, poonam.yadav\}@york.ac.uk}
   \IEEEauthorrefmark{1}School of Computer Science \& Electronic Engineering, University of Essex, Colchester, UK\\ Email: Syed.Shah@essex.ac.uk
   }
}



\maketitle

\begin{abstract}
The demands of ultra-reliable low-latency communication (URLLC) in ``NextG" cellular networks necessitate innovative approaches for efficient resource utilization. The current literature on 6G O-RAN primarily addresses improved mobile broadband (eMBB) performance or URLLC latency optimization individually, often neglecting the intricate balance required to optimize both simultaneously under practical constraints. This paper addresses this gap by proposing a DRL-based resource allocation framework integrated with meta-learning to manage eMBB and URLLC services adaptively. Our approach efficiently allocates heterogeneous network resources, aiming to maximize energy efficiency (EE) while minimizing URLLC latency, even under varying environmental conditions. We highlight the critical importance of accurately estimating the traffic distribution flow in the multi-connectivity (MC) scenario, as its uncertainty can significantly degrade EE. The proposed framework demonstrates superior adaptability across different path loss models, outperforming traditional methods and paving the way for more resilient and efficient 6G networks.
\end{abstract}

\begin{IEEEkeywords}
eMBB, DRL, URLLC, Resource Allocation, O-RAN.
\end{IEEEkeywords}

\IEEEpeerreviewmaketitle

\vspace{-0.25cm}
\section{Introduction}
The relentless evolution of wireless communication technologies has paved the way for unprecedented connectivity, catalyzing transformative changes across various sectors of society. From the advent of 1G analogue systems to the current era dominated by 5G networks, each generation has witnessed remarkable advancements in terms of data rates, latency, and network capacity \cite{bs1}\cite{bgg1}. However, as we stand on the cusp of the sixth generation (6G) of cellular networks, the paradigm is set to shift once again, ushering in an era characterised by ultra-reliable low-latency communication (URLLC), massive machine-type communication (mMTC), and enhanced mobile broadband (eMBB) services \cite{bff1}\cite{6GG}.
6G networks, often referred to as ``NextG" networks, are envisioned to be the cornerstone of the future digital infrastructure, empowering a wide variety of emerging technologies such as autonomous vehicles, augmented reality, and the Internet of Things (IoT) \cite{6GAI}\cite{ws}. 
The essence of 6G lies in its ability to provide seamless connectivity with ultra-low latency and high reliability, particularly for URLLC applications in a highly energy-efficient manner \cite{6GURLLC}. These applications are critical in scenarios where real-time communication and immediate feedback are paramount. However, achieving stringent URLLC requirements, such as latency below one millisecond and reliability above 99.999\%, poses significant challenges, particularly when coupled with the need for improved energy efficiency (EE) \cite{RanaR4}.
EE is an increasingly vital consideration in the design and deployment of NextG networks. The proliferation of connected devices, the growing demand for data-intensive applications, and the environmental concerns associated with high energy consumption necessitate the development of energy-efficient network architectures.
To realize the full potential of NextG networks, it is imperative to design intelligent algorithms and network architectures that meet stringent performance requirements and ensure EE, scalability and flexibility \cite{bcg}. 
In this context, the Open Radio Access Network (O-RAN) concept has garnered significant attention as a promising framework for building agile and cost-effective cellular networks \cite{b1}\cite{b11}. 
O-RAN introduces disaggregation and virtualization of network components, enabling interoperability between hardware and software from different vendors. This disaggregated architecture promotes innovation and facilitates the deployment of various services customized to specific user requirements \cite{Oran}\cite{Oran2}. 

Within the realm of 6G O-RAN, one of the key challenges is the efficient allocation of network radio resources to meet the diverse needs of eMBB and URLLC applications \cite{bnn1} while maximizing overall network EE.
Traditional resource allocation strategies do not consider network EE and often optimize either overall network throughput, eMBB throughput, or URLLC latency individually, overlooking the intricate trade-offs between the two. Moreover, the heterogeneity of traffic demands and the dynamic nature of wireless channels exacerbates the complexity of resource allocation in 6G networks \cite{PhyCom}\cite{bvv1}. Addressing these challenges requires novel approaches that leverage advanced techniques such as multi-connectivity (MC) and network slicing (NS) to steer traffic intelligently across the network. MC architecture refers to the design and implementation of network systems that enable users to simultaneously connect to multiple types of network or multiple network layers, thus improving reliability, capacity, and coverage. This architecture is crucial in developing NextG networks, where seamless connectivity and high performance are mandatory \cite{3GPPMC}\cite{3GPPDC}\cite{3GPPMC2}.
Energy-efficient resource allocation has been widely studied in the literature; however, most of these works mainly focus on optimizing transmit power and beamforming vectors \cite{EEPower1, EEPower2}. The authors in \cite{EERadio1} have proposed an RRM algorithm that maximizes DL EE while minimizing interference and enhancing network performance in heterogeneous wireless networks based on cognitive radio. Their simulation results show improved QoS, fairness, and EE. The work in \cite{EERadioCloud} focuses on enhancing EE in heterogeneous cloud RAN through joint optimisation of resource block (RB) assignment and power allocation via intelligent cell association and interference mitigation. More specifically, it presents an enhanced soft fractional frequency reuse scheme to allocate radio resources effectively, distinguishing between high- and low-rate-constrained QoS requirements. The optimisation problem is addressed using an equivalent convex feasibility problem and an iterative algorithm. The proposed approach results in significant EE improvements in heterogeneous cloud RAN over traditional architectures.
An energy-efficient radio resource management (RRM) approach for wireless networks is proposed in \cite{ExplainAI}. This study tackles the performance-explainability trade-off in AI models using Kernel SHAP, CERTIFAI, and Anchors methods to generate feature importance explanations and simplify a reinforcement learning (RL) agent. The goal of the RL agent is to learn optimal RRM decisions to reduce network energy consumption. The methodology reduces the RL agent's complexity by 27-62\% without losing performance. It shows that using an Anchors-based inference process can replace an AI-based process with similar performance but higher interpretability.

In URLLC and eMBB enabled NextG networks, ML-based RRM has garnered significant research attention due to its potential to optimize network performance by balancing stringent low latency and high reliability requirements with high data throughput needs \cite{RanaCM}. In \cite{RanaICC}, the authors address resource allocation challenges in multicell wireless systems that serve both eMBB and URLLC users. They propose a distributed learning framework leveraging a Thompson sampling-based Deep Reinforcement Learning (DRL) algorithm to make real-time resource allocation decisions in O-RAN architectures. By deploying trained execution agents at Near-Real Time Radio Access Network Intelligent Controllers (Near-RT RICs) at network edges, the approach ensures efficient online decision-making. Simulation results demonstrate the effectiveness of this algorithm in meeting QoS requirements for both eMBB and URLLC users, optimising resource utilisation in dynamic wireless environments.
Likewise, \cite{RanaICCSPA} present a novel approach to energy-efficient resource allocation for eMBB and URLLC users within O-RAN environments. Utilising on-policy and off-policy transfer learning strategies within a DRL model, the proposed framework dynamically allocates RBs, makes radio resource puncturing decisions, and adjusts transmit power to optimize EE. With the help of simulation results, it is shown that the method effectively achieves rapid convergence to optimal resource distribution policies that enhance EE even under varying and unpredictable channel conditions. An energy-efficient packet delivery mechanism that incorporates frequency hopping for uplink (UL) and proactive drop for DL is proposed in \cite{JointUL}. This mechanism reduces the probability of DL outage and controls the overall reliability of DL. An optimization problem is formulated to minimize the average total power consumption under URLLC constraints, solving it with a three-step method that involves bandwidth allocation, antenna configuration, and subchannel assignment. 
Simulation results validate the approach’s effectiveness in achieving significant power savings, improving EE, and meeting URLLC QoS requirements.

In \cite{EERA1}, the authors propose a resource allocation scheme to improve EE in networks supporting eMBB and URLLC. They propose an energy-efficient and spectral-efficient non-orthogonal slicing scheme that allows eMBB and URLLC services to coexist on the same physical infrastructure. By jointly optimising beamforming and RRU selection, the study tackles the EE maximisation problem using different methods such as the Dinkelbach algorithm, reweighted \(\ell_1\)-norm, and difference of convex programming. Their simulation results show that the proposed method improves EE and optimizes resource allocation, making it a promising solution for NextG networks. In  \cite{EERA2}, the authors propose an energy-efficient resource allocation scheme for ultra-dense networks that accommodate both eMBB and URLLC users. 
They formulate the problem as a non-convex combinatorial integer fractional programming problem, which they decompose into two sub-problems: RB allocation and power allocation. The proposed alternating optimisation algorithm effectively balances EE and QoS requirements by employing continuous convex approximation and difference of convex programming. The simulation results show that the proposed algorithm significantly improves EE while ensuring QoS for both eMBB and URLLC users.

To address the issue of URLLC and eMBB resource allocation in NextG networks with MC, Li et al. \cite{MC1} have proposed a spectrum-saving approach for URLLC-enabled industrial automation by focusing on joint DL and DL transmission in a single-cell setting. Their approach integrates MC with grant-free contention-based access, data replication, and broadcasting to enhance reliability and minimise bandwidth usage. 
 The study derives the packet loss probability by considering both collisions due to contention-based access and decoding errors due to dynamic wireless channels. They analyse the optimal configuration of subcarrier spacing and transmission time intervals to support data replication within limited bandwidth and end-to-end latency constraints. In addition, they prove the relationship between collision probability and packet loss probability, identifying the optimal block length that minimizes the required bandwidth. Through simulation, it is shown that the proposed framework optimizes transmission durations, block lengths, and replica numbers to reduce bandwidth, validating the effectiveness of their cross-layer resource allocation strategy. From a perspective of vehicular networks-based MC, Xue et al. in \cite{MC2} have proposed a power allocation scheme for URLLC users that improves reliability and efficiency. They introduce a DL URLLC transmission method, where URLLC packets are duplicated and sent over multiple independent links simultaneously. This approach ensures that packet transmission fails only if all independent links fail, thus improving reliability. The power allocation problem is formulated as a multi-agent DRL framework, with each link acting as an agent, allowing dynamic adaptation to varying link numbers. The authors employ a transformer neural network architecture to enable information sharing among agents and design a cooperative multi-agent DRL algorithm, named transformer-associated proximal policy optimization, for robust power allocation with imperfect CSI. The simulation results demonstrate that the proposed approach improves URLLC reliability and EE in dynamic vehicular environments.
 
\begin{figure}[t]
    \centering
   \includegraphics[width=0.5\textwidth]{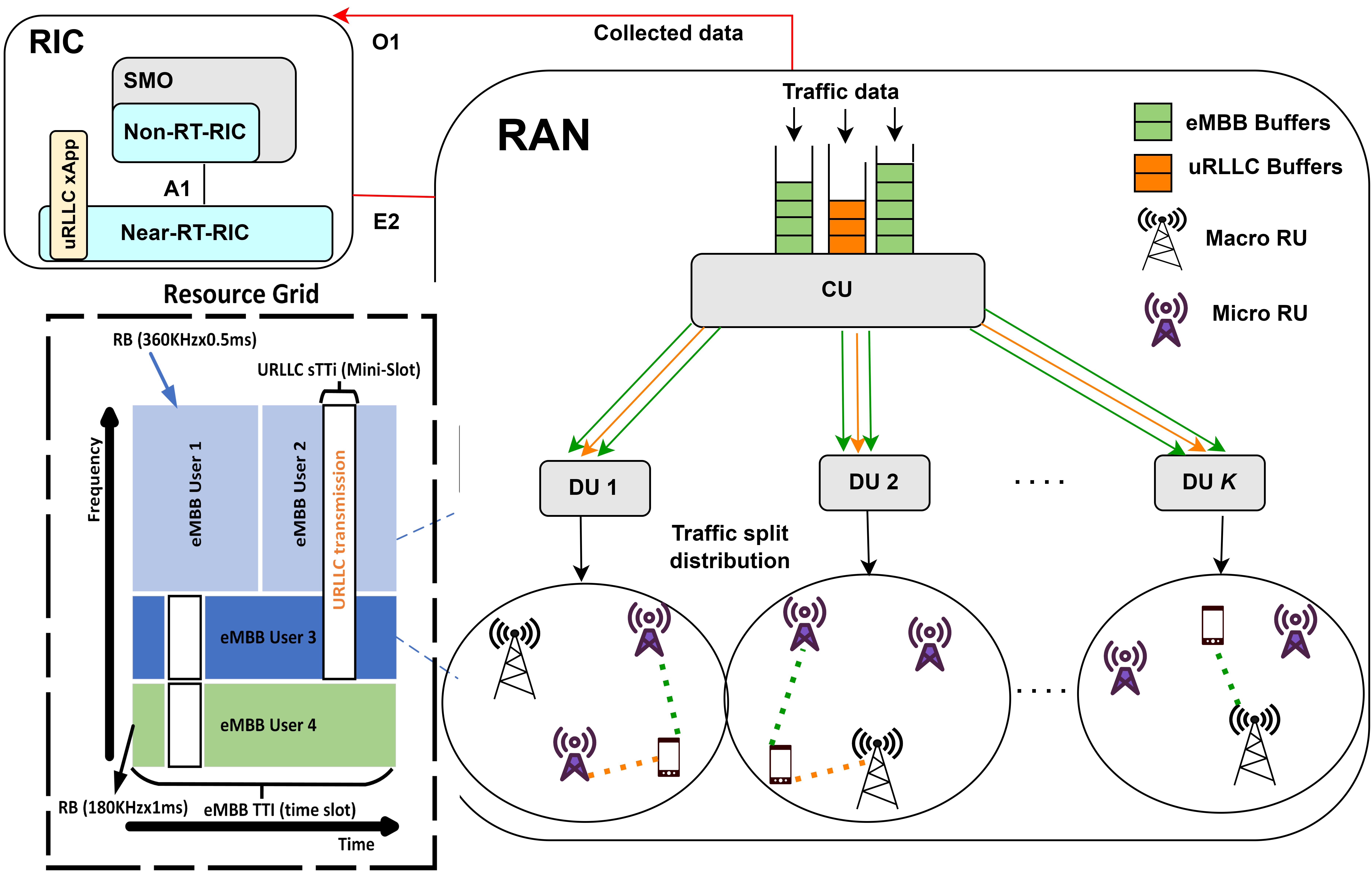}
    \caption{Multiconectivity enabled O-RAN based Small Cell Network and Considered Resource Grid.}
    \label{fig1l}
\end{figure}
\subsection{Main Contribution}
This paper proposes a comprehensive framework for energy-efficient resource utilisation in 6G O-RAN, specifically focusing on traffic steering based on MC-based joint resource scheduling techniques. By formulating the RAN resource allocation problem, we aim to maximise EE and minimise URLLC latency simultaneously, subject to stringent quality of service (QoS) requirements and practical constraints such as orthogonality, power efficiency and limited front-haul capacity.
This paper contributes several contributions to the existing 6G O-RAN resource management literature.
Unlike previous studies that predominantly address the optimisation of eMBB or URLLC in isolation, our proposed framework offers a holistic approach that considers the interaction between these two key performance metrics. By jointly optimising EE and URLLC latency, we strive to achieve a balanced allocation of network resources that cater to diverse application requirements.
We leverage the MC-based joint resource scheduling approach to enable efficient traffic steering across the network. Our approach dynamically allocates resources based on real-time traffic demands and channel conditions. 
The key contributions are as follows:
\begin{itemize}
    \item We propose an intelligent resource allocation framework based on Proximal Policy Optimization (PPO), designed to jointly optimise the EE and URLLC latency in 6G O-RAN environments. This approach efficiently balances the demands of both the eMBB and URLLC services, addressing the limitations of existing methods focusing on either aspect in isolation.
    \item The proposed framework is enhanced with both on-policy and off-policy meta-learning strategies in O-RAN, enabling adaptive and real-time resource management across varying path loss models and network environments. This integration significantly improves the system's performance in terms of latency and energy efficiency, especially under dynamic network conditions.
    \item The study emphasises the critical role of accurate traffic distribution flow estimation between the eMBB and URLLC services. We show that inaccuracies in this estimate can lead to significant degradation in EE, underscoring the need for precise traffic management in heterogeneous network environments.
    \item We perform a thorough performance evaluation of the proposed framework in different path loss models, including urban, rural, and indoor environments. 
\end{itemize}
\begin{figure*}[t]
  \centering
  \includegraphics[width=\textwidth]{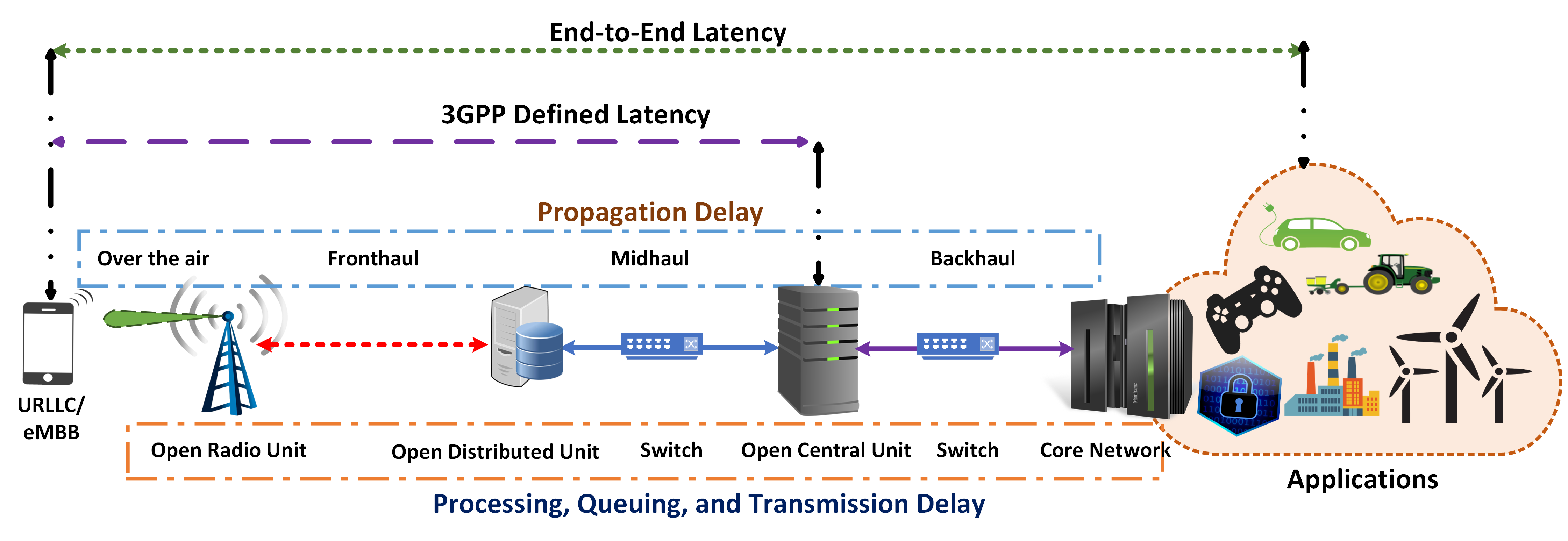}
  \caption{Different delays in O-RAN based Small Cell Network.}
  \label{fig2}
\end{figure*}
\vspace{-0.1cm}

\section{System Model}
We consider a DL orthogonal frequency-division multiple access (OFDMA) system within the ORAN framework, featuring $K$ number of DUs $\mathcal{K}=\{1, 2, . . ., K \}$, served by one CU. 
We assume that $k^{th}$ DU serves a $N_{k}$ number of multi-antenna RUs, denoted as $\mathcal{N}_{k}=\{1, 2, . . ., n, . . ., N_{k}\}$, which consists of both macro RU and micro RU. Each DU $k\in K$ serves the sets of eMBB users $\mathcal{U}_{n,k}^{e}= \{1, . . ., U_{n,k}^{e} \}$, and URLLC users $\mathcal{U}_{n,k}^{ur}= \{1, . . ., U_{n,k}^{ur} \}$. Each DU is connected to the Near-RT-RIC, which features a URLLC xApp responsible for scheduling URLLC users intelligently, denoted by \( \overline{U} = \{1, 2, \ldots, U\} \), where \( U = \{U_{n,k}^{e} + U_{n,k}^{ur}\} \) pertains to their respective RUs as shown in Fig. \ref{fig1l}.
Micro RUs are uniquely capable of managing high-density data, making them the perfect option for meeting the demands of URLLC services. Conversely, macro RUs offer high data rates and broad coverage for eMBB users. 
In the MC configuration, the operating frequency of the macro RU is independent of the micro RU frequencies. Let $\varpi_0$ and $\varpi_1$ represent the sets of sub-band frequencies used by the macro RU and micro RUs, respectively.
The allocation of radio spectrum in 5G New Radio (5GNR) can be visualised across frequency and time domains. These domains are subdivided into smaller segments known as RBs, constituting a sum of $M$ radio resources. 
Each frame is segmented into different time slots, and each time slot is further subdivided into $L$ mini-slots, with each having a duration equal to one transmission-time interval (TTI). Each RB is characterised by a distinct bandwidth, denoted as $\nabla$. Generally, the eMBB service utilises multiple TTIs to improve spectral efficiency (SE). However, the strict latency requirements mandate the prompt processing of incoming URLLC traffic.
To satisfy the strict demands for URLLC traffic, the technique of puncturing eMBB slots is employed. Dedicating certain resources to URLLC traffic through this method guarantees timely and reliable delivery of critical information. The URLLC service is designed with a short TTI of 0.25 ms, in contrast to the longer 1 ms duration allocated for eMBB service. Rapid URLLC transmission, which involves pausing eMBB traffic, can considerably affect the system's capacity and reliability. As a result, the efficiency of eMBB services might decline. Therefore, an appropriate framework is essential to ensure QoS standards are met. Our objective is to enhance the performance of the RAN in three dimensions: time, frequency, and power. The challenge involves allocating the total number of ($M= \varpi_{0} + \varpi_{1}$) RBs to its users across all $N$ RUs managed by the given $k$ DU.
\subsection{Throughput Estimation}
The binary variable $\phi_{m,l}^{n,u}(t)\in\{0,1\} $ represents the puncturing decision. Here,  $\phi_{m,l}^{n,u}(t) = 1$ when the $u^{th}$ URLLC user punctures the $l^{th}$ mini-slot, and $\phi_{m,l}^{n,u}(t) = 0$ otherwise. This is valid for every $u$ in the set $\overline{U}$ and $n$ in the set $\mathcal{N}$ at time $t$. The achievable rate for an eMBB user $w$ in RU $n$, using RB $m$ at time slot $t$, can be expressed as 
\begin{align}
    r_{n,m}^{e,u}(t)=\nabla\left(1-\frac{\sum_{l=1}^{L}\phi_{m,l}^{n,u}(t)}{L} \right)\log_{2}\big(1+\zeta_{n,m}^{e,u}(t) \big),
\end{align}
where the term $\frac{\sum_{l=1}^{L}\phi_{m,l}^{n,u}(t)}{L}$ represents the reduction in the eMBB rate caused by puncturing, and $\zeta_{n,m}^{e,u}(t)$ refers to the signal-to-noise-and-interference-ratio (SINR) of the eMBB user and it can be expressed as follows
\begin{equation}
    \zeta_{n,m}^{e,u}(t)=\frac{p_{n,m}^{e,u}(t)g_{n,m}^{e,u}(t)}{\!\!\!\!\sum\limits_{\substack{n'\in\mathcal{N}_k \\ n'\neq n}}\underbrace{p_{n',m}^{e,u}(t)g_{n',m}^{e,u}(t)}_{\text{eMBB interference}}+\!\!\!\!\sum\limits_{\substack{n'\in\mathcal{N}_k \\ n'\neq n}}\underbrace{p_{n',m}^{ur,u}(t)g_{n',m}^{ur,u}(t)}_{\text{URLLC interference}}+\sigma^{2}},
\end{equation}
where $p_{n,m}^{e,u}(t)$ and $g_{n,m}^{e,u}(t)$ represent the transmitted power and channel gain, respectively, for eMBB user $u$ associated with RU $n$ on RB $m$. Additionally, $\sigma^{2}$ refers to the noise power.
We assume that each RU assigns an RB to a single user. We introduce a binary decision variable \(\Theta_{u,m}^{n,k}(t)\in\{0,1\}\) for RB allocation. Here, \(\Theta_{u,m}^{n,k}(t)=1\) if RB \(m\) from RU \(n\) in DU \(k\) is allocated to the eMBB user \(u\), and \(\Theta_{u,m}^{n,k}(t)=0\) otherwise. This is true for all \(n\) in the set \(\mathcal{N}\) and \(k\) in the set \(\mathcal{K}\). Thus, the overall throughput obtained by the eMBB user $u$ can be determined as 
\begin{align}
     r_{n,k}^{e,u}(t)=\sum\limits_{m\in\mathcal{M}}\Theta_{u,m}^{n,k}(t)r_{n,m}^{e,u}(t).
\end{align}

To ensure fast and reliable communication in URLLC systems, it is essential to limit the size of data packets. This is because Shannon’s capacity theorem, which applies when large block lengths, does not directly address scenarios with finite block lengths \cite{bs2}. The throughput for URLLC with a finite block length can be calculated as follows
\begin{align}
     r_{n,m}^{ur,u}(t)=&\sum\limits_{m\in\mathcal{M}}\nabla_{m}\Big(\frac{\sum_{l=1}^{L}\phi_{m,l}^{n,u}(t)}{L} \Big)\Bigg[\log_{2}\big(1+\zeta_{n,m}^{ur,u}(t) \big)\\ 
     &-\sqrt{\!\frac{\!\Lambda_{n,m}^{ur,u}}{\!\varrho_{n,m}^{ur,u}(t)}}.Q^{\!-1}(\!x)\Bigg],\nonumber
\end{align}
where $\varrho_{n,m}^{ur,u}(t)$ denotes the number of symbols per mini-slot and $\Lambda_{n,m}^{ur,u}=1-\frac{1}{(1+\zeta_{n,m}^{ur,u}(t))^{2}}$ represents the channel dispersion. Here, $\zeta_{n,m}^{ur,u}(t)$ signifies the SINR for the URLLC user, which can be expressed as
\begin{equation}
    \zeta_{n,m}^{ur,u}(t)=\frac{p_{n,m}^{ur,u}(t)g_{n,m}^{ur,u}(t)}{\!\!\!\!\sum\limits_{\substack{n'\in\mathcal{N}_k \\ n'\neq n}}\underbrace{p_{n',m}^{ur,u}(t)g_{n',m}^{ur,u}(t)}_{\text{URLLC interference}}+\!\!\!\!\sum\limits_{\substack{n'\in\mathcal{N}_k \\ n'\neq n}}\!\!\underbrace{p_{n',m}^{e,u}(t)g_{n',m}^{e,u}(t)}_{\text{eMBB interference}}+\sigma^{2}}.
\end{equation}

\subsection{URLLC Latency Estimation}
The primary goal for URLLC users is to minimise their end-to-end (e2e) latency from the CU to the end users. To meet these rigorous requirements, URLLC xApp must ensure that data is processed and transmitted without the delay introduced by queueing. This means that as soon as data arrive, it is immediately handled and sent to its destination without any waiting time, which is essential for maintaining the service's reliability and responsiveness. 
Consequently, the e2e latency experienced by URLLC users is primarily determined by the time it takes for data to be processed and transmitted through the network, excluding any delays that might be caused by queuing.
URLLC users' packet is handled at the CU layer before being routed to Virtual Network Functions (VNFs) in the DU layer for simultaneous processing. To manage the processing of these packets, we employ an M/M/1 queue model, which operates on a first-come, first-served basis. This means that packets are processed in the order they arrive. Specifically, for URLLC traffic, which follows the FTP3 model standardised by 3GPP and involves packets of $\digamma$ bytes in size, the arrival of packets is described by a Poisson process with a mean arrival rate $\Upsilon_{u}(t)$, measured in packets per second. Packets cannot be segmented and must be fully transmitted by each allocated RB.

Let $\omega_{c}$ and $\omega_{d}$ represent the processing capacity of CU and DU measured in cycles per second. For packets of the same size $\digamma$, the processing capacity required to process a single packet is indicated as $\psi$ (in cycles). Thus, the task processing rates at the CU and DU, \( \xi_{c} \) and \( \xi_{d} \), can be defined as \( \xi_{c} = \frac{\omega_{c}}{\psi} \) and \( \xi_{d} = \frac{\omega_{d}}{\psi} \), respectively. Consequently, the average service times for the CU and DU are given by \( \frac{1}{\xi_{c}} \) and \( \frac{1}{\xi_{d}} \).
The computational delay for the URLLC arrival packet at the CU (\(\lambda^{\text{pro}}_{cu}\)) and the DU (\(\lambda^{\text{pro}}_{du}\)) can be determined as 
$\lambda^{\text{pro}}_{cu}(t) = \frac{1}{\xi_{\text{c}} - \delta(t)}$ and $\lambda^{\text{pro}}_{du}(t) = \frac{1}{\xi_{\text{d}} - \delta(t)}$, respectively. 
The expression $\delta(t)=\sum_{u\in\mathcal{U}^{u}}\Upsilon_{u}(t)$ refers to the cumulative arrival rate of the URLLC packet in the CU. We presuppose that $\xi_{c}>\delta(t)$ and $\xi_{d}>\delta(t)$ ensure the stability of the queue.
Incoming packets $\Upsilon_u(t)$ for URLLC traffic are sent to the DU through the mid-haul link (MH), which has a maximum capacity $\Omega^{MH}$ (bits/sec). In particular, the average arrival data rate at the DU is nearly the same as the average arrival data rate at the initial layer.
Therefore, the transmission delay for the URLLC traffic, considering the constraints of the MH capacity, can be expressed as
\begin{align}
    \lambda_{cu,du}^{tx}(t)=\frac{\delta(t).\digamma}{\Omega^{MH}}
\end{align}
By employing the MC method, the URLLC traffic generated per TTI is divided into multiple segments. These segments are sent through different links and subsequently combined at the user's end.
The maximum $N_{k}$ number of routes can be utilised to deliver data from the DU $k$ to a user, allowing flexibility. 
We define the vector for the traffic distribution for the user \( u \) by \( \vartheta_u(t) = [\vartheta_{0,u}(t), \vartheta_{1,u}(t), \vartheta_{2,u}(t), \ldots, \vartheta_{N_k,u}(t)] \), where \( \sum_{n \in \mathcal{N}_k} \vartheta_{n,u}(t) = 1 \). Here, \( \vartheta_{n,u} \) represents the portion of traffic routed to user \( u \) through RU \( n \).
When sending data packets to a particular user, these packets can be transmitted using multiple RUs. The goal is to determine the effective response time $\lambda_{du,ru}^{tx}$ for these packets to travel through the DU to the user.
To find this effective response time, we look at the response times of the different front-haul (FH) links connected to the DU. Each FH link has a certain maximum data transmission capacity $\Omega^{FH}_{n}$. However, since we want to ensure all packets are delivered efficiently, we base our calculation on the slowest average response time among these FH links, such as
\begin{align}
    \lambda_{du,ru}^{tx}(t)=\max_{n}\left\{\frac{\sum_{u\in\mathcal{U}_{n,k}^{ur}}\vartheta_{n,u}(t)\Upsilon_{u}(t)\digamma}{\Omega^{FH}_{n}} \right\}, \forall n\in\mathcal{N}_k  
\end{align}
The delay from RU to a user is subsequently determined as
\begin{align}
    \lambda_{ru,u}^{tx}(t)=\max_{n}\left\{\frac{\vartheta_{n,u}(t)\Upsilon_{u}(t)\digamma}{r_{n,m}^{ur,u}(t)} \right\}, \forall u\in\mathcal{U}_{n,k}^{ur}
\end{align}
Thus, the e2e latency of URLLC user per TTI can be determined as follows
\begin{equation}
\begin{split}
    \lambda_{u}^{ur}(t)=\lambda^{\text{pro}}_{cu}(t)+\lambda^{\text{pro}}_{du}(t)+ \lambda_{cu,du}^{tx}(t)+\lambda_{du,ru}^{tx}(t)\\ + \lambda_{ru,u}^{tx}(t)+\lambda^{\text{pro}}_{ru}(t),\hspace{0.3cm} \forall u\in\mathcal{U}_{n,k}^{ur} 
    \end{split}
\end{equation}


\section{Problem Formulation}
The main objective of this paper is to jointly optimize traffic steering and energy-efficient resource utilization of the eMBB and URLLC service in 6G O-RAN, subject to different constraints.
We define the EE of the system by determining the proportion of the total data rate to the overall power usage, specifically
\begin{align}
   \gamma_{EE}^{k}(t)\!= \!\frac{\sum_{n\in\mathcal{N}_k}\sum_{u\in \mathcal{\Bar{U}}}\{r_{n,k}^{e,u}(t) + r_{n,m}^{ur,u}(t) \}}{\sum_{n\in\mathcal{N}_k}\!\sum_{u\in \mathcal{\Bar{U}}}\!\sum_{m\in\mathcal{M}}p_{n,m}^{u}(t)\! +\! N_k\!\cdot\!P_{RU}(t) \!+ \!P_{DU}^{k}},
\end{align}
where \(N_k \cdot P_{RU}(t)\) represents the total power consumed by \(N_k\) RUs, and \(P_{DU}^{n}\) denotes the power consumption of the \(n^{th}\) DU at each TTI. Accordingly, we define an objective function as follows
\begin{subequations}\label{eq:op}
 \small 
\begin{align}
\max_{\vartheta, \Theta , P, \phi}&\left\{\gamma_{EE}^{k}(t)\right\} \label{eq:obf} \\
\text{s.t.}\hspace{0.1cm}\text{C1:}&\sum_{u\in \mathcal{U}_{n,k}^{e}}\!\Theta_{u,m}^{n,k}(t)\leq1, \forall m\in\mathcal{M}, n\in\mathcal{N}_k, k\in\mathcal{K} \label{eq:c1}\\
\text{C2:}&\sum_{u\in \mathcal{U}_{n,k}^{ur}}\!\phi_{m,l}^{n,u}(t)\leq1, \forall m\in\mathcal{M}, n\in\mathcal{N}_k, k\in\mathcal{K}\label{eq:c2}\\
\text{C3:}&\hspace{0.15cm}\sum_{l\in \mathcal{L}}\!\phi_{m,l}^{n,u}(t)\leq L, \forall m\in\mathcal{M}, n\in\mathcal{N}_k\label{eq:c3}\\
\text{C4:}&\hspace{0.15cm}\lambda_{u}^{ur}\Big(\vartheta(t), \Theta(t), P(t), \phi(t)\Big)\leq \eta^{ur}, \forall u\in\mathcal{U}_{n,k}^{ur}\label{eq:c4}\\
\text{C5:}&\sum_{u\in\mathcal{U}_{n,k}^{e}}r_{n,k}^{e,u}(t)\geq \Bar{r}_{e}\label{eq:relc} \\
\text{C6:}&\sum_{u\in\mathcal{U}_{n,k}^{e}}\sum_{m\in\mathcal{M}}p_{n,m}^{e,u}(t)\leq P_{max}, \forall n\in\mathcal{N}_k\label{eq:c5}\\
\text{C7:}&\hspace{0.15cm}p_{n,m}^{e,u}(t)\geq 0, \forall u\in\mathcal{U}^{e}, m\in\mathcal{M} \label{eq:c6}\\
\text{C8:}&\hspace{0.15cm}\Theta_{u,m}^{n,k}(t)\in\{0,1\}, \forall u\in\mathcal{U}^{e}, m\in\mathcal{M}\label{eq:c7}\\
\text{C9:}&\hspace{0.15cm}\phi_{m,l}^{n,u}(t)\in\{0,1\}, \forall u\in\mathcal{U}^{ur}, m\in\mathcal{M}\label{eq:c8}\\
\text{C10:}&\hspace{0.15cm}\sum_{u}\left[r_{n,k}^{e,u}(t) + r_{n,m}^{ur,u}(t) \right]\leq\Omega^{FH}_{n}, \forall n\in\mathcal{N}_k \label{eq:c10}\\
\text{C11:}&\hspace{0.15cm}r_{n,m}^{ur,u}(t)\geq\frac{\vartheta_{n,u}(t)\Upsilon_{u}(t)\digamma}{M} \forall n\in\mathcal{N}_k, u\in \mathcal{U}^{ur} \label{eq:c11}
\end{align}
\end{subequations}
where $\vartheta$, $\Theta$, $P$, and $\phi$ refer to the traffic steering, eMBB resource allocation, eMBB power allocation and URLLC resource scheduling decision variables, respectively. Here, C1 denotes the limit on allocating eMBB resources, ensuring that each RB is allocated to only one user. Constraint C2 refers to the fact that only one URLLC user can puncture a particular mini-slot within an RB. Constraint C3 specifies that the number of punctured mini-slots should not exceed the overall number of available mini-slots. Constraint C4 guarantees that the minimum latency requirement for the URLLC user is met, with end-to-end latency limited by a set threshold. Constraint C5 indicates the
reliability of the eMBB. Constraints C6 and C7 define the power allocation limits for eMBB. Similarly, constraints C8 and C9 specify the restrictions related to the allocation of resources. Constraint C10 denote the FH capacity limitations between DU and RU. Lastly, constraint C11 guarantees that every RB allocated to the URLLC user must transmit an entire data packet of size $\digamma$.

\section{Proposed Intelligent Framework}
The objective function defines the optimization problem, and the constraints C1 through C11 are characterized by their combinatorial and binary nature. Constraints C8 and C9 involve binary decision variables, which typically contribute to the problem's complexity. Furthermore, constraints related to resource allocation and power limits introduce additional combinatorial challenges. 
This structure indicates that the optimization problem in (\ref{eq:op}) is NP-hard. Therefore, due to its NP-hard nature, we anticipate that exact solutions may be impractical for large-scale instances of the problem. Consequently, heuristic and DRL-assisted frameworks are used to obtain feasible solutions within a reasonable timeframe. 
\subsection{Heuristic Approach for optimizing the $\vartheta(t)$}
We employ xApp1 on near-RT-RIC, which is based on a heuristic method to estimate the traffic distribution decision $\vartheta(t)$ for traffic stream separation. We propose a dynamic approach for traffic split decision-making in Open RAN that incorporates EE and adapts the dynamic window size based on the Channel Quality Indicator (CQI).
This method provides a responsive and efficient approach to traffic split decisions by adapting to varying network conditions.
First, we collect the CQI values for each user \( u \) and RU \( n \) over the past \( J \) time slots:
\begin{align}
    \text{CQI}_{n,u}[t-J+1], \text{CQI}_{n,u}[t-J+2], \ldots, \text{CQI}_{n,u}[t].
\end{align}
Next, we compute the mean $\overline{\text{CQI}}$ of these CQI values as follows 
\begin{align}
\overline{\text{CQI}}[t] = \frac{1}{J} \sum_{i=t-J+1}^{t} \text{CQI}_{n,u}[i],
\end{align}
 
To determine the appropriate window size, we compute the autocorrelation function (ACF) of the collected CQI values over the past $J$ time slots. The ACF helps us to understand the temporal correlation of CQI values, which is critical for dynamically adjusting the window size. 
High autocorrelation values indicate stable channel conditions for certain periods, suggesting that a larger window size may be beneficial. Low autocorrelation values indicate rapidly changing conditions, suggesting a need for a shorter window size to capture recent variations accurately.
The autocorrelation function $\Phi(z)$ at lag $z$ for the CQI values is given by
\begin{align}
    \Phi(z) = \frac{\sum_{i=1}^{J-z} (\text{CQI}_{n,u}[i] - \overline{\text{CQI}}) (\text{CQI}_{n,u}[i+z] - \overline{\text{CQI}})}{\sum_{i=1}^{J} (\text{CQI}_{n,u}[i] - \overline{\text{CQI}})^2}
\end{align}
The dynamic window size \( \Gamma \) is determined based on the computed ACF. Specifically, we identify the delay at which the ACF drops below a specified threshold \( \aleph \). This lag represents the window size that captures the significant temporal dependencies in the CQI values while ignoring the less relevant distant dependencies.
\begin{align}
\Gamma = \min\left\{z : \Phi(z) < \aleph\right\}
\end{align}
The calculated window size \( \Gamma \) is then ensured to be within the predefined optimal range \( [\Gamma_{\text{min}}, \Gamma_{\text{max}}] \):

\begin{align}
\Gamma = \max\left(\Gamma_{\text{min}}, \min\left(\Gamma_{\text{max}}, \Gamma\right)\right)
\end{align}
where \( \Gamma_{\text{min}} \) and \( \Gamma_{\text{max}} \) are the minimum and maximum window sizes.
Using the dynamic window size \( \Gamma \), we compute the EE for RU $n$ and user $u$ for each traffic type $x$ at time $t$ over a window size $\Gamma$ as 
\begin{align}
EE_{m,u}^x[t] = \frac{\sum_{i=t-\Gamma + 1}^{t}{R_{n,u}^x[i]}}{\sum_{i=t-\Gamma + 1}^{t}{P_{n}^x[i]}}.
\end{align}
To smooth the energy efficiency values, we apply a weighted moving average:
\begin{align}
\overline{EE_{n,u}^x[t]} = \frac{\sum_{i=t-\Gamma+1}^{t} \mu_{i} EE_{n,u}^x[i]}{\sum_{i=t-\Gamma+1}^{t} \mu_{i}},
\end{align}
where \( \mu_{i} \) is the weight for time frame \( i \). For exponential weights, \( \mu_{i} = \alpha^{t-i} \), where \( \alpha \) is a decay factor.

Finally, the traffic distribution ratio \( \hat{\vartheta}_{n,u}^x[t] \) is determined by normalising the weighted moving average of EE as follows
\begin{align}
\hat{\vartheta}_{n,u}^x[t] = \frac{\overline{EE_{n,u}^x[t]}}{\sum_{n} \overline{EE_{n,u}^x[t]}}.
\end{align}
This method ensures that the traffic distribution ratio is optimized both for performance and energy efficiency, leading to a more balanced and effective network operation.
\subsection{DRL-based Resource Scheduling}
The objective of the DRL-based framework is to optimize the assignment of RBs to users in a way that maximizes the overall EE of the network while adhering to the limitations. This involves framing the problem as a Markov decision process (MDP) that includes $N_k$ agents, each with its own state and actions but sharing a common goal. In a multi-agent framework, every agent functions within its distinct state and action domains. The state domain includes the particular segment of environmental observations available to each agent, while the action domain consists of the unique set of actions that each agent can perform.
The separation of state and action spaces facilitates a more organized and efficient approach for agents to perform their tasks and cooperate. As a result, they can effectively work together towards achieving a unified objective.
\subsubsection*{State space}
We define the state of the \( i \)-th DRL agent in DU \( k \) as \( s_{i,k} \). The complete set of states for all \( N_k \) DRL agents in DU \( k \) is then given by $ \mathcal{S}_k = \{ s_{1,k}, s_{2,k}, \ldots, s_{n,k},\ldots, s_{N_{k},k} \},$ where \( \mathcal{S}_k \) represents the set of states of all DRL agents in DU \( k \). The state space of each agent consists of the channel conditions of the eMBB and URLLC users, incoming traffic information, and estimated traffic flow-split distribution at time slot $t$. It can be represented as $s_{n,k}(t)=\{g_{n,k}^{e}(t), g_{n,k}^{ur}(t), \Upsilon_{n,k}(t), \vartheta_{n,k}(t), U_{n,k}^{e}, U_{n,k}^{ur} \}$.   
\subsubsection*{Action}
The set of actions for all \( N_k \) agents in DU \( k \) is \( \mathcal{A}_k = \{ a_{1,k}, a_{2,k}, \ldots, a_{N_k,k} \} \).
Each DRL agent takes actions regarding the eMBB RBs allocation \( \Theta(t) \), eMBB power allocation \( P(t) \), and URLLC scheduling \( \phi(t) \). 
\subsubsection*{Reward}
To achieve energy-efficient resource utilisation in 6G O-RAN systems, we define the global reward function \( R(t) \) for each agent at time \( t \) as follows: 

\begin{align}
\begin{split}
R(t) = \gamma_{EE}^{k}(t) - \upsilon_1 \cdot \sum_{u \in \mathcal{U}_{n,k}^{ur}} \max\left(0, \lambda_{u}^{ur}(t) - \eta^{ur}\right)- \\ \upsilon_2 \cdot \max\left(0, \bar{r}_e - \sum_{u \in \mathcal{U}_{n,k}^{e}} r_{n,k}^{e,u}(t)\right),
\end{split}
\end{align}
where \( \gamma_{EE}^{k}(t) \) represents the energy efficiency metric calculated as the ratio of the total data rate to the overall power usage. The first penalty term accounts for the URLLC latency, penalising the reward if the latency exceeds the threshold \( \eta^{ur} \). The weights \( \upsilon_1 \) and \( \upsilon_2 \) allow for fine-tuning the emphasis on URLLC latency and eMBB reliability relative to optimizing energy efficiency.
The second penalty term penalises the reward if the eMBB data rate falls below the required threshold \( \bar{r}_e \), ensuring reliability for eMBB users. 
\subsection{PPO-based approach}
A fundamental aspect of DRL is the development of an optimal policy that effectively maps the state of the network to actions. Given that the action space includes both discrete and continuous values, the conventional Deep Q-Networks (DQN) method cannot be directly used for decision making, as it yields discrete outputs. However, this discretisation not only expands the action space's dimensionality but also introduces quantisation errors \cite{rn}. 
Proximal Policy Optimization (PPO) offers a robust framework to address the complex problem of RB allocation. PPO is a policy gradient method designed to optimize the policy in RL scenarios, particularly when dealing with large, continuous action spaces. The PPO algorithm balances exploration and exploitation, employing a clipped objective function, which ensures that policy updates are not too drastic, thus enhancing stability and convergence.
In PPO, the policy \( \pi_{\theta}(a|s) \) is parameterised by \( \theta \), and the objective is to find the optimal policy parameters \( \theta \) that maximize the expected cumulative reward. The core of the PPO algorithm involves optimizing the surrogate objective function, which is defined as follows:

\begin{align}
\gimel^{\text{PPO}}(\theta) = \mathbb{E}_t \left[ \min \left( r_t(\theta) \hat{A}_t, \text{clip}(r_t(\theta), 1 - \epsilon, 1 + \epsilon) \hat{A}_t \right) \right],
\end{align}
were \( r_t(\theta) \) is the probability ratio, given by:
\begin{align}
r_t(\theta) = \frac{\pi_{\theta}(a_t|s_t)}{\pi_{\theta_{\text{old}}}(a_t|s_t)},
\end{align}
where \( \pi_{\theta_{\text{old}}} \) represents the policy parameters from the previous iteration. The advantage function \( \hat{A}_t \) estimates the relative advantage of taking action \( a_t \) at state \( s_t \). The clipping mechanism, controlled by the hyperparameter \( \epsilon \), prevents large policy updates, ensuring that the new policy does not deviate significantly from the old policy. This clipping helps maintain stable training and prevents large, destructive policy updates that could degrade performance. The advantage estimates \( \hat{A}_t \) can be calculated using Generalized Advantage Estimation (GAE), which allows more stable and efficient updates. GAE combines the benefits of bootstrapping from the value function and Monte Carlo methods, offering a trade-off between bias and variance in the advantage estimates. The temporal difference error \( \beta_t \) is computed as:
\begin{equation}
\beta_t = r_t + \chi V_\Psi(s_{t+1}) - V_\Psi(s_t)
\end{equation}
where \( V_\Psi(s_t) \) is the estimated value function. The GAE can then be defined as:
\begin{equation}
\hat{A}_t = \sum_{i=0}^{\infty} (\chi \daleth)^i \beta_{t+i},
\end{equation}
where $\chi\in \{0,1\}$ refers to the discount factor and $\daleth$ is the GAE parameter that determines the trade-off between bias and variance. 
The policy parameters $\theta$ are then updated by maximizing the clipped surrogate objective function $\gimel^{PPO}(\theta)$. The update employs stochastic gradient ascent method, and the gradient is calculated with respect to the parameter $\theta$ as follows
\begin{align}
    \theta' = \theta + \alpha\Delta\gimel^{PPO}(\theta)
\end{align}
where $\alpha$ is the learning rate. The policy is updated over multiple iterations, allowing it to learn from the experience collected and refine its actions over time.
The value function \( V_\Psi(s_t) \) is used to estimate the expected return from the state, and it is updated to minimise the mean squared error loss as follows
\begin{align}
    \gimel^{VF}(\Psi) = \mathbb{E}_t \left[ \left( V_\Psi(s_t) - R_{t}\right)^2\right],
\end{align}
where $R_t$ is the actual return at time $t$.
In our specific problem, the PPO algorithm's reward function \( R(t) \) is designed to capture the trade-off between energy efficiency, eMBB reliability, and URLLC latency. 
Using PPO, we ensure that policy updates are both efficient and stable, which is crucial given the channel environment's dynamic and complex nature. The algorithm's ability to handle large action spaces and continuous adjustments aligns well with resource allocation and scheduling requirements. Moreover, the clipping mechanism in PPO effectively addresses the challenge of maintaining a balance between exploration and exploitation, which is critical in a multi-agent setup with shared goals and constraints.
\begin{algorithm}[t]
\caption{PPO for Resource Allocation}
\label{alg:ppo}
\begin{algorithmic}[1]
\STATE Initialize policy parameters \( \theta \), value function parameters \( \Psi \)
\STATE Set hyperparameters \( \epsilon \), \( \chi \), and \( \daleth \)

\FOR{each episode}
    \STATE Initialize the environment and state \( s_0 \)
    \FOR{ \( t = 0 \) to \( T \)}
        \STATE Sample action \( a_t \) from the policy \( \pi_{\theta}(a_t|s_t) \)
        \STATE Execute action \( a_t \), observe reward \( r_t \) and next state \( s_{t+1} \)
        \STATE Store transition \( (s_t, a_t, r_t, s_{t+1}) \)
    \ENDFOR
\ENDFOR

\FOR{each time step \( t \)}
    \STATE Compute the temporal difference error according to (23)
    \STATE Compute advantages using (24)
\ENDFOR

\FOR{each update iteration}
    \STATE Compute the surrogate objective using (21)
    \STATE Update the policy parameters according to (25)
\ENDFOR

\STATE Update value function parameters using (26)

\STATE Repeat steps 1-5 for a fixed number of episodes or until convergence.
\end{algorithmic}
\end{algorithm}
\section{Adaptive scheduling via Meta-Learning Framework }
In the previous section, we introduced a DRL-based method to tackle the issues of eMBB RB allocation, power distribution, and URLLC scheduling in a static environment. The PPO-based framework assumes that the environment remains consistent during training and testing. However, this assumption does not hold true in a real-time wireless communications environment. Current research falls short in terms of generalization to varied wireless channel conditions because differences can occur if the testing environment's conditions differ from those of the training environment. We propose an adaptive meta-learning (AML) approach that enhances generalization to previously unseen scenarios to address this issue.
The actor update in this framework relies on two foundational concepts: on-policy meta-learning for RB allocation at the near-RT-RIC and off-policy meta-learning for the non-RT-RIC.
By integrating PPO with on-policy meta-learning at the near-RT-RIC and off-policy meta-learning at the non-RT-RIC, we can achieve a comprehensive and adaptive resource allocation framework for 6G O-RAN. Near-RT-RIC focuses on rapid adaptation and real-time optimization, while non-RT-RIC handles strategic, long-term optimization tasks. This combined approach ensures that the system can respond to immediate changes in network conditions while maintaining a robust and optimized policy over time.
\subsection{Near-RT-RIC with On-Policy Meta-Learning}
The near-RT-RIC operates within a tight latency budget, typically on the order of milliseconds to seconds. The near-RT-RIC makes rapid and adaptive decisions to optimize resource allocation dynamically.
On-policy meta-learning is particularly suited for this environment because it can quickly adapt to new tasks with limited data by leveraging experiences gathered from similar tasks. We employ the Model-Agnostic Meta-Learning (MAML) approach for On-policy meta-learning, which enhances the adaptability of PPO by enabling rapid adaptation to new and unseen tasks with minimal data \cite{MAML}.

To effectively implement on-policy meta-learning in the near-RT-RIC, defining a distribution of tasks the system is expected to encounter is crucial. Each task corresponds to a unique set of network conditions, such as varying user demands, channel quality, and resource availability. The goal is to enable the near-RT-RIC to generalize well across these tasks, allowing efficient and effective resource allocation.
The initialisation of the policy parameters is a critical step in the meta-learning process. The near-RT-RIC seeks to find an optimal set of policy parameters $\theta$ that can quickly adapt to new tasks. The meta-learning process begins with initialising these parameters, which can be achieved by training on a diverse set of tasks. The objective during this training phase is to minimize the cumulative loss across all tasks, expressed as:
\begin{equation}
\gimel_{meta}(\theta) = \sum_{i=1}^{C} \gimel_{task_i}(\theta),
\end{equation}
where $\gimel_{task_i}(\theta)$ represents the loss function for task $i$ and $C$ is the number of tasks.
\subsubsection*{Meta-Training Phase}
During the meta-training phase, the near-RT-RIC collects data from multiple tasks. For each task $i$, the near-RT-RIC optimises the PPO objective function defined in (21). For each task, the near-RT-RIC updates the policy parameters using the gradients obtained from the PPO objective. The update rule for task $i$ can be expressed as:
\begin{equation}
\theta_i' = \theta - \alpha \Delta_\theta \gimel_{PPO}^i(\theta),
\end{equation}
where $\alpha$ is the learning rate, this update allows the policy parameters to adapt to the specific characteristics of task $i$.
\subsubsection*{Meta-Update}
After performing task-specific updates, the next step is aggregating the gradients from multiple tasks to update the meta-policy parameters. The meta-update is performed to refine the initialisation of the policy parameters, enabling the near-RT-RIC to generalise better across different tasks. This can be mathematically represented as:
\begin{equation}
\theta = \theta - \hat{\alpha} \sum_{i=1}^{C} \Delta_\theta \gimel_{PPO}^i(\theta_i'),
\end{equation}
where $\hat{\alpha}$ is the meta-learning rate. This aggregation of gradients ensures that the updated policy parameters reflect information from all tasks, thus enhancing the adaptability of the near-RT-RIC.
\subsubsection*{Adaptation Phase}
Once the meta-training is complete, the near-RT-RIC enters the adaptation phase, where it must quickly fine-tune the policy parameters using new data from the current task. The objective during this phase is to maximise the expected cumulative reward, which can be represented as
\begin{equation}
R = \sum_{t=0}^{T} \wp^t R(t),
\end{equation}
where $R(t)$ is the reward received at time $t$ and $\wp$ is the discount factor that values immediate rewards higher than future rewards.
During adaptation, the near-RT-RIC utilises the learned policy initialisation $\theta$ and further updates the parameters based on the new experiences from the current task. The PPO updates during this phase follow the same structure as the meta-training phase but with the current task's data:
\begin{equation}
\theta' = \theta - \alpha \Delta_\theta \gimel_{PPO}(\theta)
\end{equation}
This rapid adaptation ensures that the near-RT-RIC can effectively respond to changing network conditions in real-time.
\subsection{Non-RT-RIC with Off-Policy Meta-Learning}
The non-RT-RIC operates on a longer timescale, typically in seconds to minutes, and is responsible for more strategic and less latency-sensitive tasks, such as long-term network optimisation and policy updates. Off-policy meta-learning is well-suited for this environment because it can leverage large datasets collected over extended periods, allowing for more comprehensive learning and adaptation.
Experience replay is a fundamental technique used in off-policy learning that involves storing past experiences in a replay buffer. This buffer accumulates state-action-reward-next state tuples \((s, a, r, s')\) from interactions with the environment over time.
The replay buffer \( \mathcal{D} \) maintains a large collection of these experiences, enabling the non-RT-RIC to sample batches of past experiences and update the policy. The experiences are sampled according to their relevance and recency, ensuring that the learning process remains efficient and stable. The buffer helps mitigate issues like correlation between consecutive experiences and ensures a more stable learning process by breaking the temporal correlation.
\subsubsection*{Task Sampling and Meta-Training Phase}
In off-policy meta-learning, the non-RT-RIC aims to learn a policy that can efficiently adapt to various tasks or network conditions. The meta-learning process begins with the non-RT-RIC collecting diverse tasks and experiences over time.
For each sampled task \(i\), the PPO algorithm optimises the policy by updating the policy parameters \( \theta \). 
The policy parameters \( \theta \) are updated using the gradient of the PPO objective function with respect to \( \theta \) as follows

\begin{equation}
\theta_i' \leftarrow \theta - \alpha \Delta_\theta \gimel_{PPO}^i(\theta).
\end{equation}
This process involves optimising the policy based on the experiences collected from the current task, which helps adapt the policy to the task's specific requirements.
\subsubsection*{Meta-Update}
After updating the policy parameters for each task, the non-RT-RIC performs a meta-update to improve the policy's generalisation across multiple tasks. This step aggregates the policy gradients obtained from different tasks to refine the global policy. The meta-update can be represented as
\begin{equation}
\theta \leftarrow \theta - \hat{\alpha} \sum_i \Delta_\theta \gimel_{PPO}^i(\theta_i').
\end{equation}
This aggregation helps combine the knowledge gained from various tasks, leading to a more robust policy adaptable to different network conditions.
\subsubsection*{Off-Policy Updates}
Incorporating off-policy updates involves using experiences stored in the replay buffer to update the policy. The non-RT-RIC samples a batch of experiences from the buffer and performs updates based on this off-policy data. This approach allows the non-RT-RIC to use a larger dataset, improving the sample efficiency of the learning process.
The policy update using off-policy data can be expressed as
\begin{equation}
\small
\begin{split}
\theta \leftarrow \theta - \eta \Delta_\theta\!\! \left[\! \frac{1}{|B|}\!\!\sum_{\substack{(s, a, r, s') \\ \in B}}\!\!\!\!\!\left( r + \wp V_\Psi(s') - V_\Psi(s) \right) \Delta_\theta \log \pi_\theta(a | s) \right],
\end{split}
\end{equation}
where \( B \) is a batch of experiences sampled from the replay buffer, \( \eta \) is the learning rate, and \( V_\Psi(s) \) is the value function approximator. This update uses off-policy data to adjust policy parameters \( \theta \) and improve policy performance across different tasks.
\begin{algorithm}[t]
\caption{Adaptive scheduling using Meta-learning}
\begin{algorithmic}[1]
\STATE Initialize meta-policy parameters $\theta$ and value function parameters $\phi$,  meta-learning rates $\alpha$ (inner loop) and $\hat{\alpha}$ (outer loop). 
\STATE Initialize replay buffer $\mathcal{B}$ for off-policy learning
\FOR{each meta-iteration}
    \STATE Sample batch of tasks $\{T_i\}$ from task distribution $p(T)$
    \FOR{each task $T_i$}
        \IF{on-policy meta-learning}
            \STATE Collect trajectories $\{\tau_i\}$ using policy $\pi_\theta$ on $T_i$
        \ELSE
            \STATE Sample trajectories $\{\tau_i\}$ from replay buffer $\mathcal{B}$
        \ENDIF
        \STATE Compute rewards-to-go $\hat{R}_t$ and advantages $\hat{A}_t$ using GAE
        \STATE Perform policy update using Algorithm 1 (PPO)
        \STATE Store collected trajectories in replay buffer $\mathcal{B}$ (if off-policy)
    \ENDFOR
    \STATE Perform meta-update using (33)
\ENDFOR
\end{algorithmic}
\end{algorithm}
\subsection{Complexity Analysis}
The proposed framework of on-policy meta-learning at the near-RT-RIC and off-policy meta-learning at the non-RT-RIC involves a combination of computationally intensive processes, resulting in a nuanced complexity profile. The on-policy meta-learning aspect at the near-RT-RIC entails frequent updates due to the real-time nature of the environment, with the complexity dominated by the PPO updates, which are \(O(C \cdot T \cdot A \cdot Q)\), where \(C\) is the number of tasks, \(T\) is the number of time steps per task, \(A\) is the action space dimension, and \(Q\) is the number of policy network parameters. The adaptation phase, where rapid fine-tuning occurs, contributes an additional \(O(Y \cdot P)\) per adaptation step, with \(Y\) representing the number of meta-updates and \(P\) the parameters in the policy. In the non-RT-RIC, off-policy meta-learning leverages experience replay buffers, where the complexity is primarily influenced by the size of the buffer and the number of off-policy updates, leading to \(O(B \cdot O)\), with \(B\) as the buffer size and \(O\) as the number of off-policy updates. Aggregating gradients across multiple tasks introduces additional computational load, approximated by \(O(C \cdot G)\), where \(G\) is the complexity of gradient calculation for each task. Consequently, the overall complexity of the combined framework is a function of these intertwined processes, reflecting the trade-off between real-time adaptability and long-term optimisation efficiency.

\section{Simulation Results}
In our simulation, we implemented the PPO algorithm with both on-policy and off-policy meta-learning strategies. We used a network model consisting of a DU that forms a set of one macro RU with three sectors, each comprising one small RU.  
The coverage area of small RUs is 100 m. All users are evenly and randomly distributed within a region of interest with a radius of 500 meters. The traffic arrival process is modeled using the Poisson process distribution for both URLLC and eMBB.  
The policy network consisted of three fully connected layers.
Each of these layers used the ReLU (Rectified Linear Unit) activation function, which is defined as \( \text{ReLU}(x) = \max(0, x) \). 
Similarly, the value network consists of three fully connected layers and has the same structure as the policy network.
The training process involved optimizing the PPO objective using the Adam optimizer, with a learning rate of \(3 \times 10^{-4}\). The PPO clipping parameter \(\epsilon\) was set to 0.2 to ensure stable policy updates. Furthermore, the discount factor \(\chi\) was set at 0.99, balancing considering immediate and future rewards. In contrast, the GAE parameter was set to 0.95 to manage the bias and variance trade-off in the advantage estimates.
Table I provides the simulation parameters. We set up the simulation setup according to the 3GPP TR 38.901 guidelines \cite{3GPPCM}.
We trained our DRL model using the Urban Macro path loss model based on the 3GPP TR 38.901 standard and considered this as a baseline, which provides guidelines for channel models for 5G NR. The specific parameters used for this model are derived from the urban environment to reflect dense urban deployments with high user density.

\begin{table}[t]
    \centering
    \caption{Simulation Configurations}
    \label{tab:simulation_config}
    \scriptsize
    \begin{tabular}{|l|c|c|c|}
        \hline
        \textbf{Properties} & \textbf{Urban} & \textbf{Rural} & \textbf{Indoor} \\ \hline
        RU Antenna model & \begin{tabular}[c]{@{}c@{}}15 dB Cosine, \\ 65\textdegree{} HPBW\end{tabular} & \begin{tabular}[c]{@{}c@{}}15 dB Cosine, \\ 65\textdegree{} HPBW\end{tabular} & \begin{tabular}[c]{@{}c@{}}15 dB Cosine, \\ 40\textdegree{} HPBW\end{tabular} \\ \hline
        Pathloss model & 3GPP UMa & 3GPP RMa & 3GPP InH \\ \hline
        Macro RU Transmit power & \multicolumn{3}{c|}{40 dBm} \\ \hline
        Small RU Transmit power & \multicolumn{3}{c|}{26 dBm} \\ \hline
        BW of Macro RU & \multicolumn{3}{c|}{20 MHz} \\ \hline
        BW of small RU & \multicolumn{3}{c|}{100 MHz} \\ \hline
        Noise figure & \multicolumn{3}{c|}{RU: 9 dB, UE: 5 dB} \\ \hline
        URLLC packet length & \multicolumn{3}{c|}{32 bytes} \\ \hline
        Frame duration & \multicolumn{3}{c|}{10 ms} \\ \hline
        No. of OFDM symbols/TTI & \multicolumn{3}{c|}{14} \\ \hline
        No. of minislots/TTI & \multicolumn{3}{c|}{7} \\ \hline
        OFDM symbols/mini-slot & \multicolumn{3}{c|}{2} \\ \hline
        Subcarrier spacing & \multicolumn{3}{c|}{15 KHz} \\ \hline
    \end{tabular}
    \normalsize
\end{table}
    
\begin{itemize}
    \item Urban (UMa) environment: This environment characterises the propagation conditions typical of dense urban areas. It accounts for factors such as high building density and line-of-sight (LOS) propagation. 
    \item Rural (RMa) environment: This model represents scenarios with moderate building density and incorporates both LoS and NLoS conditions. 
    \item Indoor (InH) environment: This model simulates the propagation characteristics within indoor environments, where walls and furniture significantly affect signal strength, leading to predominantly NLoS conditions.
\end{itemize}
\subsection{Rural environment}
\begin{figure}[t]
    \centering
   \includegraphics[width=0.48\textwidth]{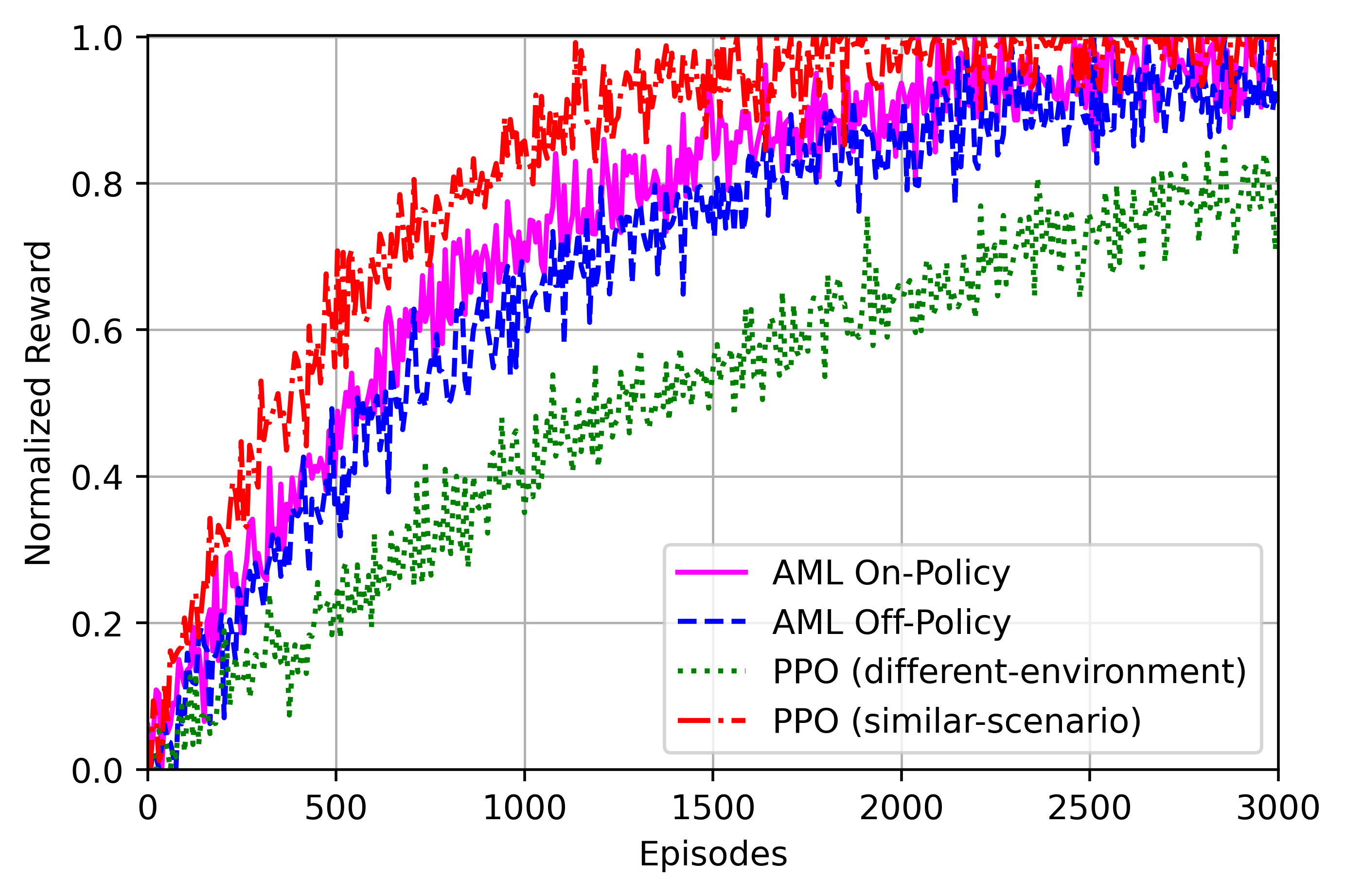}
    \caption{Convergence performance}
    \label{fig1con}
\end{figure}

Initially, we analyzed the results in the rural environment for the
proposed schemes, where we train the model in the
baseline environment (i.e., urban environment) and test it in the
rural environment.
It is known that the agent's performance improves with increased experience in varying channel conditions. However, accumulating this experience requires a significant amount of time. Therefore, we have introduced the AML approach (algorithm 2) to optimize learning in fewer steps, with the AML model being trained and tested in varying channel environments. The results illustrated in Fig. \ref{fig1con} indicate that the PPO approach (similar scenario) serves as a performance upper bound, as the model experiences consistent channel conditions during both training and testing. The uniformity in the path loss characteristics allows the agent to develop a highly specialised policy, optimizing resource allocation and power management efficiently. 
The AML on-policy approach, which adapts the policy in near-real-time at near RT-RIC, shows excellent performance. It updates the policy based on the most recent experiences, ensuring that the agent is always adapting to the current state of the environment. This leads to more relevant and contextually appropriate updates, allowing the agent to adapt rapidly to changing network conditions and traffic patterns, leading to more effective resource allocation decisions in real time. In contrast, it can be observed that Off-policy performed reasonably well, but it was generally outperformed by the on-policy approach. The use of a replay buffer allowed the off-policy agent to learn from a broader range of experiences; however, it lacked the same level of adaptation to real-time conditions. Consequently, the inability to quickly adapt to changing conditions can lead to suboptimal scheduling and power management decisions. When we evaluated the PPO (different environment), the performance decreased significantly because the policies learned in the urban setting may not generalize well to the rural scenario, resulting in inefficient resource allocation and increased latency for URLLC users. 
This highlights the importance of timely and relevant feedback in reinforcement learning for URLLC resource scheduling.

\begin{figure}[t]
    \centering
   \includegraphics[width=0.48\textwidth]{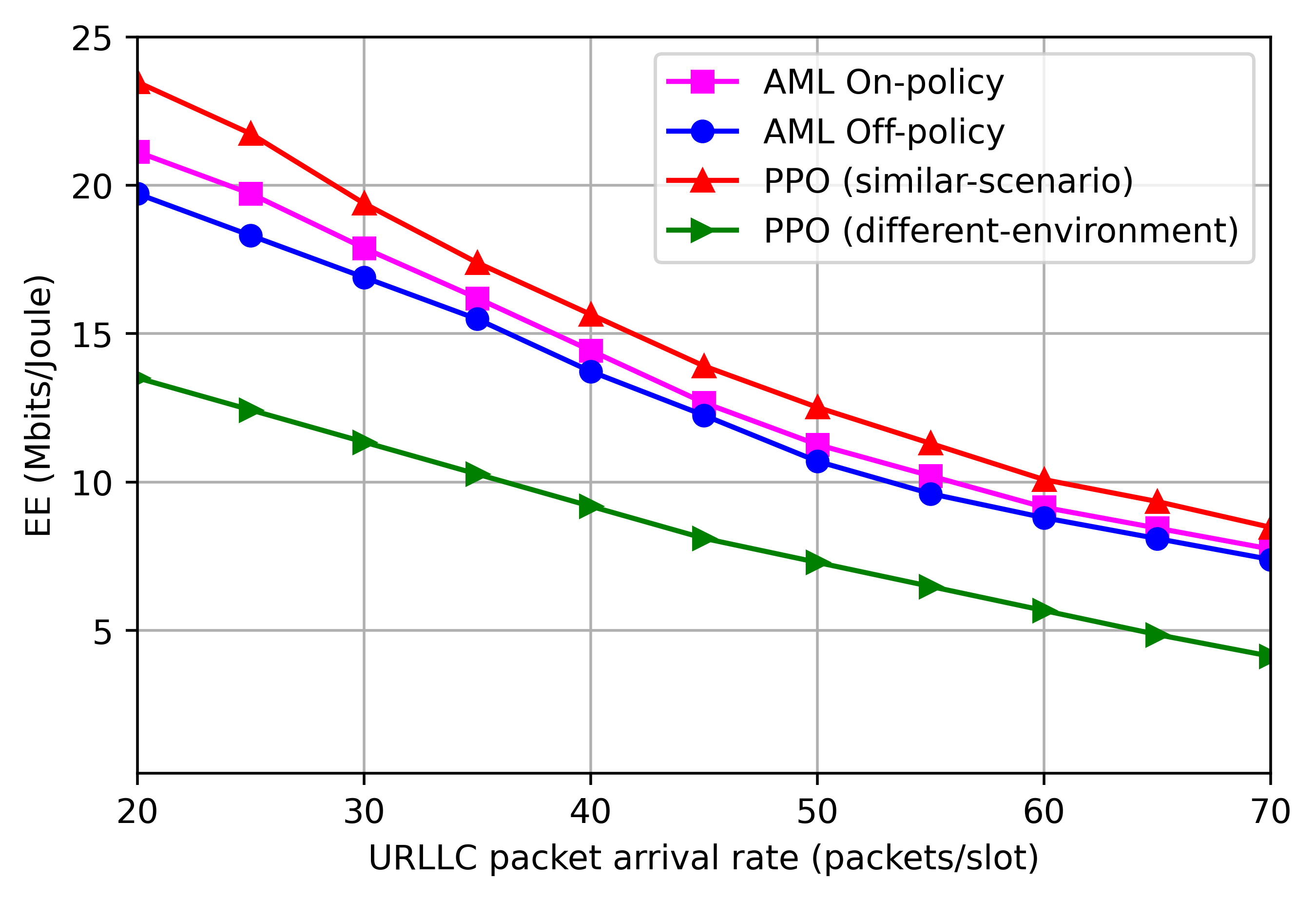}
    \caption{Impact of incoming URLLC traffic on EE of the system in rural pathloss model}
    \label{figee_r}
\end{figure}

In Fig. \ref{figee_r}, the EE performance of the system is presented under varying incoming URLLC traffic when the agent knows the flow split estimation $\vartheta$. As URLLC packet arrival rates increase, there is a noticeable decrease in energy efficiency across all approaches, reflecting the higher demand for resources to meet the stringent latency and reliability requirements of URLLC traffic. On-Policy maintains the highest EE due to its ability to adapt quickly to real-time traffic conditions, optimizing resource allocation effectively.
 Off-policy exhibits lower EE as it relies on outdated experiences from a replay buffer. PPO (similar scenario) provides a strong benchmark, but shows a decrease in efficiency as packet rates increase. PPO (different environment) experiences the most significant decline in EE, highlighting the challenges of policy generalization across different environments. This illustrates the effectiveness of policy methods in managing resource allocation efficiently under varying traffic demands.
 \begin{figure}[t]
    \centering
   \includegraphics[width=0.48\textwidth]{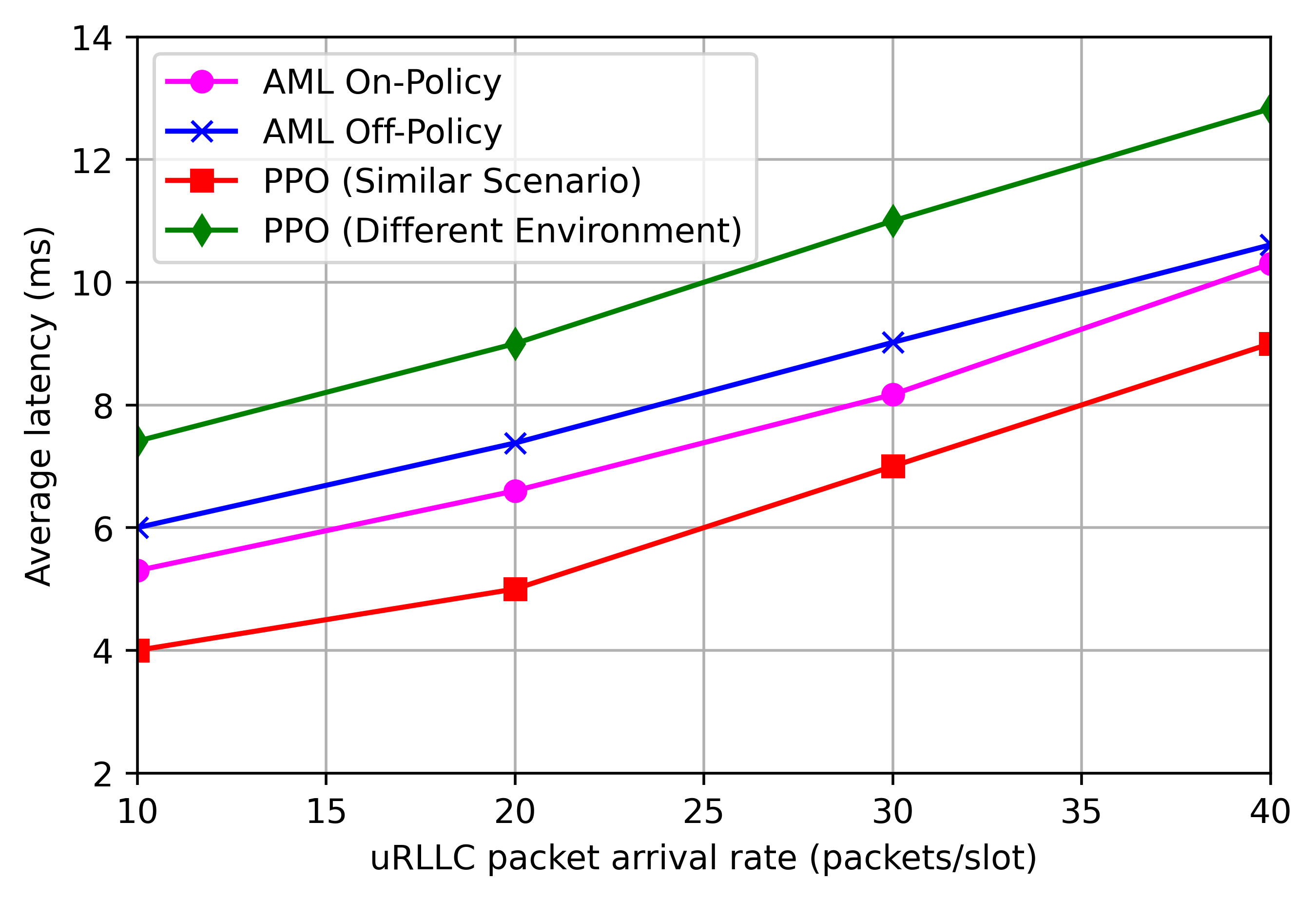}
    \caption{Average URLLC latency in rural environment}
    \label{figee_lr}
\end{figure}

In Fig. \ref{figee_lr}, we show the performance of the URLLC latency where the on-policy maintains an average latency between 5 and 15 ms, performing slightly better than off-policy, which ranges from 6 to 15 ms, due to its real-time adaptability. In general, the results highlight the critical importance of training in a consistent environment and the benefits of policy approaches in dynamic settings. 
\begin{figure}[t]
    \centering
   \includegraphics[width=0.48\textwidth]{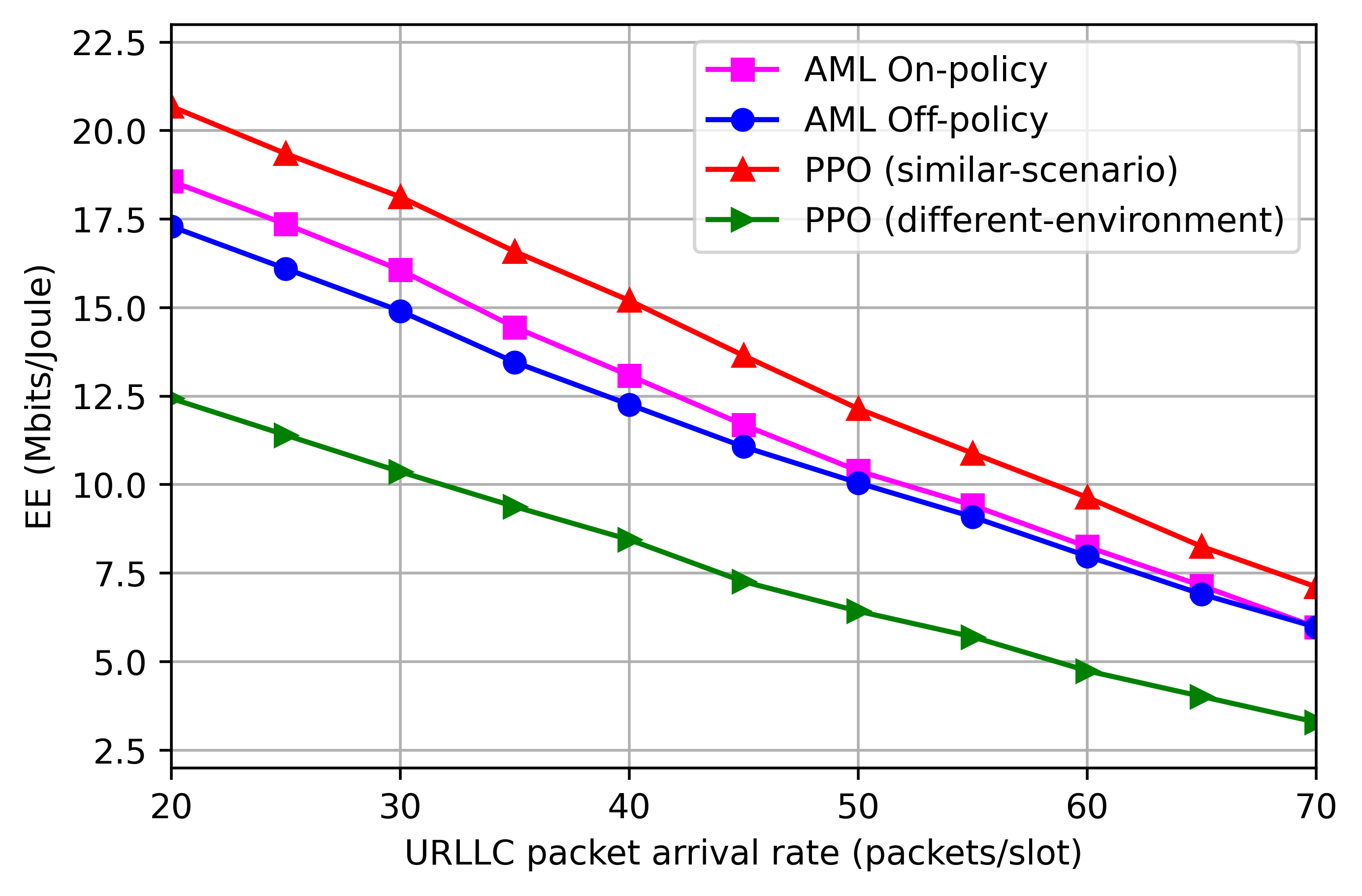}
    \caption{Impact of incoming URLLC traffic on EE of the system in rural pathloss model when $\vartheta$ is unknown}
    \label{figee_ro}
\end{figure}
Next, we present the EE performance in Fig. \ref{figee_ro} where the $\vartheta$ is unknown. It can be seen that the performance of the system in terms of EE suffers when we do not optimize the traffic split distribution compared to Fig. \ref{figee_r}. It can be due to suboptimal resource allocation, as the system lacks precise information on how traffic is distributed across users. This uncertainty can lead to inefficient power usage and improper scheduling, resulting in higher energy consumption and reduced data rates, ultimately compromising EE. 
\subsection{Indoor environment}
In the previous subsection, we explored the system's performance in the rural scenario. Now, we analyse the efficiency of the proposed approach in a new unseen channel environment (indoor).
\begin{figure}[t]
    \centering
   \includegraphics[width=0.48\textwidth]{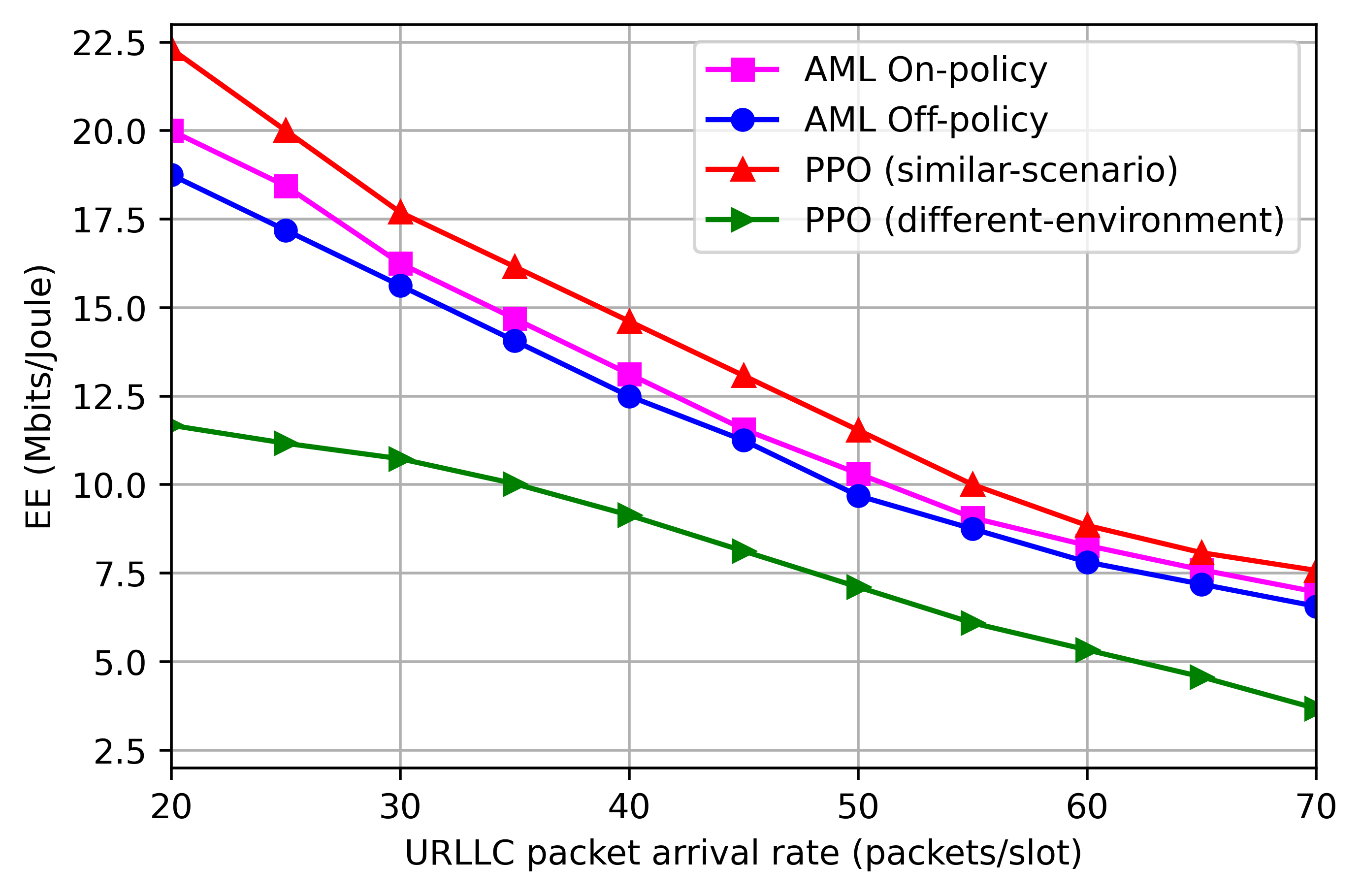}
    \caption{Impact of incoming URLLC traffic on EE of the system in indoor path loss model}
    \label{figee_ino}
\end{figure}
In Fig. \ref{figee_ino}, the EE performance in an indoor environment shows a notable degradation compared to the rural scenario. This performance drop is attributed to the distinct characteristics of indoor environments, such as severe multipath fading and shadowing effects, which differ significantly from the urban model used during training. Despite this, the AML on-policy approach continues to provide good generalization ability, maintaining better EE by dynamically adjusting to challenging indoor conditions. 
\begin{figure}[t]
    \centering
   \includegraphics[width=0.48\textwidth]{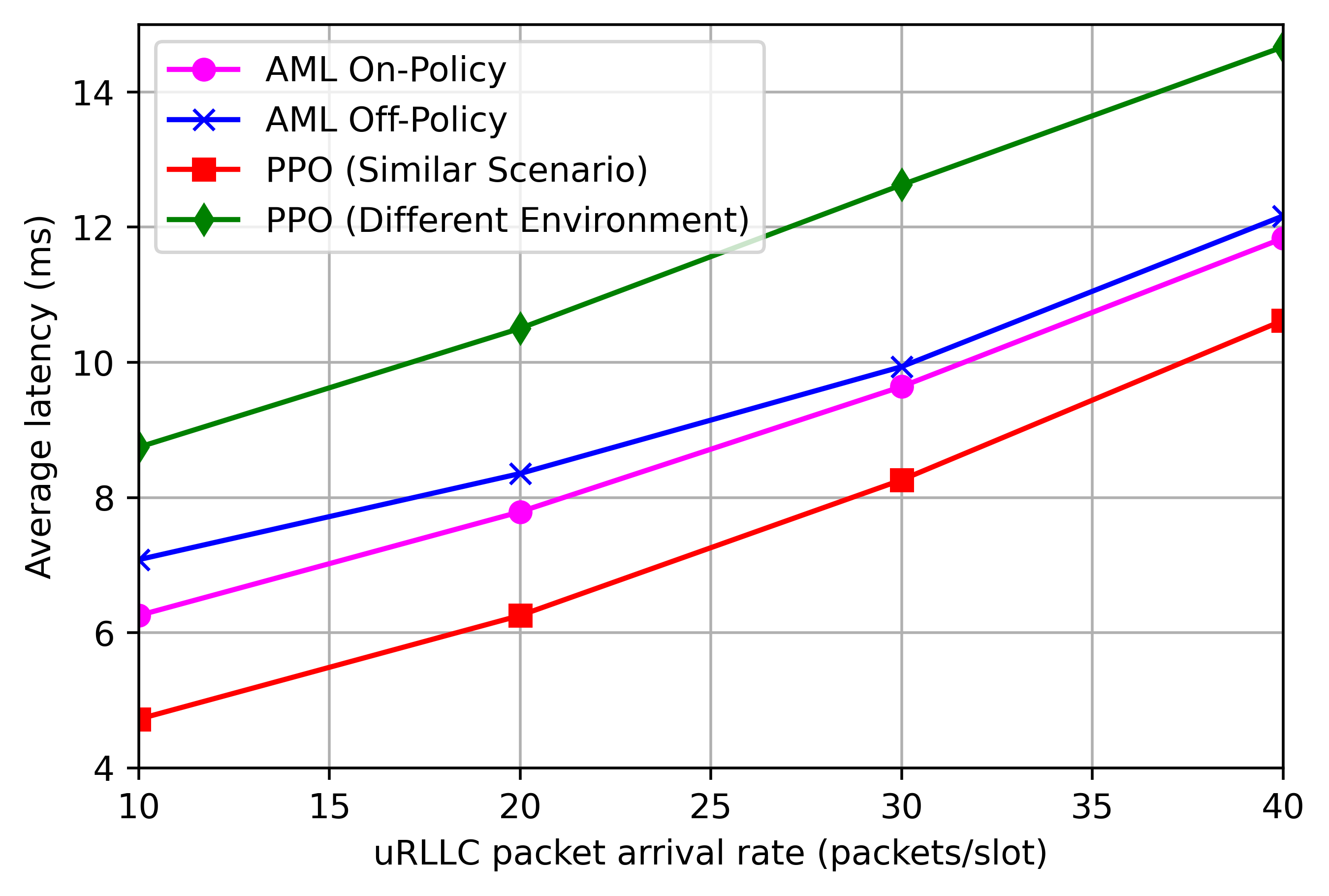}
    \caption{Average URLLC latency in indoor environment}
    \label{figl_ino}
\end{figure}

In Fig. \ref{figl_ino}, when testing in an indoor path loss model, the average latency further degrades compared to the rural scenario, reflecting the challenging indoor propagation conditions. Specifically, the latency increases more sharply with increasing URLLC packet arrival rates as the indoor environment introduces more severe multipath fading and signal attenuation. The AML on-policy approach still performs better than the off-policy and PPO (different environments), maintaining a lower latency under these complex conditions. 

\section{Conclusion}
This paper presents a comprehensive study on optimising resource allocation for eMBB and URLLC services in 6G O-RAN using DRL with both on-policy and off-policy meta-learning approaches. The results demonstrate that PPO, when trained and tested in the same environment, consistently achieves superior performance, particularly in minimizing latency, which is critical for URLLC services. However, performance degrades significantly when the model is applied in different path loss scenarios, such as rural or indoor environments, highlighting the importance of environmental consistency during training. On-policy meta-learning is more adaptable and resilient across different scenarios than off-policy methods, showcasing its potential in real-time adaptive resource management.
This study provides crucial insight into the importance of accurately estimating the traffic split flow between eMBB and URLLC services. When this variable is unknown or inaccurately estimated, EE suffers due to suboptimal resource allocation, underscoring the need for precise traffic management in dynamic and heterogeneous network environments. Future work should focus on enhancing the adaptability of learning algorithms to varying environmental conditions and improving the estimation techniques for traffic split flow to further optimise the balance between latency, reliability, and energy efficiency in 6G networks.
\section*{Acknowledgment}
This work is supported by EPSRC and DSIT through the Communications Hub for Empowering Distributed Cloud Computing Applications and Research (CHEDDAR) under grants EP/X040518/1, EP/Y037421/1 and EP/Y019229/1.
We thank Prof. David Grace, Dr. Yifan Liu and Mohit Bidikar from the University of York, UK for their various discussions on this work and helpful suggestions.

\ifCLASSOPTIONcaptionsoff
  \newpage
\fi


\bibliographystyle{IEEEtran}
\balance 
\bibliography{ref}

\newpage

\end{document}